# LOAD DISTRIBUTION COMPOSITE DESIGN PATTERN FOR GENETIC ALGORITHM-BASED AUTONOMIC COMPUTING SYSTEMS

Vishnuvardhan Mannava[1] and T. Ramesh[2]

[1]Department of Computer Science and Engineering, K L University, Vaddeswaram, 522502, A.P., India
`vishnu@kluniversity.in`
[2]Department of Computer Science and Engineering, National Institute of Technology, Warangal, 506004, A.P., India
`rmesht@nitw.ac.in`

## ABSTRACT

*Current autonomic computing systems are ad hoc solutions that are designed and implemented from the scratch. When designing software, in most cases two or more patterns are to be composed to solve a bigger problem. A composite design patterns shows a synergy that makes the composition more than just the sum of its parts which leads to ready-made software architectures. As far as we know, there are no studies on composition of design patterns for autonomic computing domain. In this paper we propose pattern-oriented software architecture for self-optimization in autonomic computing system using design patterns composition and multi objective evolutionary algorithms that software designers and/or programmers can exploit to drive their work. Main objective of the system is to reduce the load in the server by distributing the population to clients. We used Case Based Reasoning, Database Access, and Master Slave design patterns. We evaluate the effectiveness of our architecture with and without design patterns compositions. The use of composite design patterns in the architecture and quantitative measurements are presented. A simple UML class diagram is used to describe the architecture.*

## 1. INTRODUCTION

In a Genetic Algorithm (GA) application, many individuals derive, independently and concurrently, competing solutions to a problem. These solutions are then evaluated for fitness and individuals survive and reproduce based upon their fitness. Eventually, the best solutions emerge after generations of evolution [1].

The flow of a typical GA simulation is as follows [5]: First, a GA server creates many individuals randomly. Each of these individuals is tested for fitness. Based on their fitness, measured by a fitness function that quantifies the optimality of a solution, the server selects a percentage of the individuals that are allowed to crossover with each other, analogous to gene sharing through reproduction in biological organisms. The crossover between two parents produces offspring, which have a chance of being randomly mutated. A child thus produced is then placed into the population for the next generation, in which it will be evaluated for fitness. The process of selection, crossover, and mutation repeats until the new population is full and the new generation repeats the behavior of the previous generation. After many generations, the individuals are expected to become more adept at solving the problem to which the GA is being applied.

In order for a GA simulation to work well, there needs to be a significant number of individuals within a population, and the simulation needs to be allowed to run for many generations. Furthermore, the simulation will typically need to be run repeatedly while parameters such as mutation rate, population size and crossover functions are tuned. Thus, a successful GA simulation requires the calculation of the fitness function thousands of times or more. It is therefore critical that the function that performs the calculation of the fitness, called the fitness function, can be executed as speedily as possible [5].

Design Patterns have, over the last decade, fundamentally changed the way we think about the design of large software systems [6]. Using Design Patterns not only helps designers exploit the community's collective wisdom and experience as captured in the patterns, it also enables others studying the system in question to gain a deeper understanding of how the system is structured, and why it behaves in particular ways. And as the system evolves over time, the patterns used in its construction provide guidance on managing the evolution so that the system remains faithful to its original design, ensuring that the original parts and the modified parts interact as expected. Although they are not components in the standard sense of the word, patterns may, as has been noted, be the real key to reuse since they allow the reuse of design, rather than mere code.

Distributed computing applications grow in size and complexity in response to increasing computational needs, it is increasingly difficult to build a system that satisfies all requirements and design constraints that it will encounter during its lifetime. Many of these systems must operate continuously, disallowing periods of downtime while humans modify code and fine-tune the system. For instance, several studies document the severe financial penalties incurred by companies when facing problems such as data loss and data inaccessibility. As a result, it is important for applications to be able to self-reconfigure in response to changing requirements and environmental conditions. Figure 1 show the typical architecture of the autonomic computing system [5].

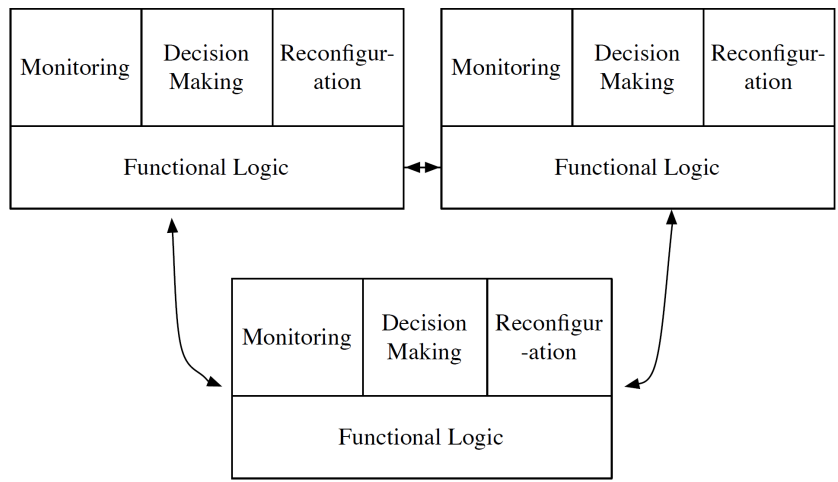

Figure 1: Autonomic computing

Proposed system solves multi object optimization using Genetic Algorithms. Multi objective optimization is a vector of decision variables which satisfies constraints and optimizes a vector function whose elements represent the objective functions. These functions form a mathematical description of performance criteria which are usually in conflict with each other. Hence, the terms "optimize" means finding such a solution which would give the values of all the objective functions acceptable to the decision maker.

The decision variables are the numerical quantities for which values are to be chosen in an optimization problem. These quantities are denoted as $x_j = 1, 2,..., n$. The vector x of n decision variables is represented by[5]:

$$X = \begin{bmatrix} X_1 \\ X_2 \\ . \\ . \\ X_n \end{bmatrix}$$

This can be written more conveniently as: $x = [x_1, x_2,...., x_n]^T$, where T indicates the transposition of the column vector to the row vector.

$g_i(x) \leq 0 \; i = 1,...,m$ or equalities:
$h_j(x) = 0 \; j = 1,...,p$

Note that p, the number of equality constraints, must be less than n, the number of decision variables, because if $p \geq n$ the problem is said to be over constrained, since there are no degrees of freedom left for optimizing (i.e., in other words, there would be more unknowns than equations). The number of degrees of freedom is given by $n - p$. Also, constraints can be explicit (i.e., given in algebraic form) or implicit, in which case the algorithm to compute $g_i(x)$ for any given vector x must be known. Proposed composite design pattern will solve multi objective optimization using Genetic Algorithm. Proposed pattern will make the server to evaluate Genetic Algorithm based on fitness function. Server generates different population which will evaluated by different client at a time, this will reduce server load. All results of clients are stored in database.

## 2. RELATED WORK

When designing software two or more patterns are to be composed to solve a bigger problem. Pattern composition has been shown as a challenge to applying design patterns in real software systems. Composite patterns represent micro architectures that when glued together could create an entire software architecture. Thus pattern composition can lead to ready-made architectures from which only instantiation would be required to build robust implementations [23]. A composite design patterns shows a synergy that makes the composition more than just the sum of its parts [6].

We can generally classify software design approaches that utilize patterns as [17]:

**Adhoc.** There is no process to guide the development and to integrate the pattern with other design artifacts.

**Systematic.** Following the: 1) *Pattern Languages*. A pattern language provides a set of patterns that solve problems in a specific domain. 2) *Development processes*. A systematic development process defines a pattern composition approach.

Current Autonomic Computing systems are ad hoc solutions that are designed and implemented from the scratch, and there are no universal standard (or well established) Software methodologies to develop [18]. There are also significant limitations to the way in which these systems are validated [19]. As a result they are generally difficult to specify, design, verify, and validate [20]. In addition, the current lack of reusable design expertise that can be leveraged from one adaptive system to another further exacerbates the problem.

The appeal of the Intelligent Machine Design (IMD) architecture [20] to autonomic computing systems is that it is closely related to the way intelligent biological systems work; it is shown to be amenable to autonomic certification. In this context, evolutionary algorithms play a major role [21].

The evaluation of an autonomic system depends on to what extent it adopts or implements the self-* properties [22]. Further, it is very difficult for a system to fully implement all the self-* properties and in many cases it becomes redundant [15]. Thus most of the autonomic systems focus on some particular properties based on their requirements and goals.

In this context, it is of paramount importance to propose software architectures, composite design patterns, patterns languages and Search Based Software Engineering (SBSE)[21] practices for the development of autonomic systems such that software designers and/or programmers can exploit to drive their work; consequently, systems will be validated. We aim to follow the systematic approach for the entire software design.

To address these problems, we have identified and proposed different Design Patterns for adaptive and Autonomic Computing systems, and this paper proposes a Pattern Oriented Software Architecture using Design Patterns compositions, Multi Objective Evolutionary Algorithms, and Service Oriented Architecture (SOA) that will work in both standalone and distributed environments.

As far as we know, there are no studies on composition of design patterns and pattern languages for autonomic computing domain [23].

We believe that this is the first time Search Based Software Engineering (SBSE) has been applied to autonomic computing: Its goal is to re-formulate software engineering problems as optimization problems that can then be attacked with computational search [21].

The proposed autonomic systems have the following properties [22]:

Self-optimizing: Automatic monitoring and control of resources to ensure the optimal functioning, the ability to automatically tune the performance of services based on online monitoring

In this section we present some works that deal with different aspects of autonomic systems and their design. Nick Burns and Mike Bradley paper[1] discuss applying Genetic Algorithm for distributing computing we take this paper is base paper here we are applying Genetic Algorithms and Design Patterns in autonomic systems. The author of the paper uses composite, singleton half-sys and half-asyn patterns for system designing. In Jason O. Hallstrom and Neelam Soundarajan[2] uses observer pattern for monitoring approach for determining whether the pattern contracts used in developing a system are respected at runtime. In Andres J. Ramirez and David B. Knoester [3] proposes applying Genetic Algorithms for decision making in autonomic computing.

In Andres J. Ramirez and David B. Knoester[4] uses Distributed Adapters Pattern (DAP) in the context of remote communication between two components for object oriented applications. In this paper uses all base paper concepts for designing new system for autonomic computing. Genetic Algorithms also have been used to design overlay multicast networks for data distribution [9]. These overlay networks must balance the competing goals of diffusing data across the network as efficiently as possible while minimizing expenses. A common approach for integrating various objectives in a Genetic Algorithm is to use a cost function that linearly combines several objectives as a weighted sum [12]. Although most of these approaches [10] achieved rapid convergence rates while producing overlay networks that satisfied the given constraints, to our knowledge, the methods were not applied at run time to address dynamic changes in the network's environment [11].

## 3 LOAD DISTRIBUTION DESIGN PATTERN TEMPLATE:

To facilitate the organization, understanding, and application of the adaptation Design Patterns, this paper uses a template similar in style to that used by Ramirez et al. [2]. Likewise, the Implementation and Sample Code fields are too application-specific for the Design Patterns presented in this paper.

### 3.1. Pattern Name:

Load Distribution Design Pattern

### 3.2. Classification:

Structural – Monitoring-Decision Making

### 3.3. Intent:

Load Distribution Design Pattern main objective is to distribute load of genetic server to different clients. Generally every problem consists more than one solution in Genetic Algorithm server solve all possible solutions. Out of all solutions server will provide best solution as the result. So Genetic Algorithm server load will increase gradually our pattern distribute Genetic Algorithm population to different clients.

### 3.4. Motivation:

Main motive of Load Distribution Design Pattern is to distribute Genetic Algorithm population to different client for solving problem. Our goal is to reduce server load by distributing Genetic Algorithm population.

### 3.5. Proposed Design Pattern Structure:

A UML class diagram for the Service Administration Pattern can be found in Figure 2.

Three Design Patterns are used for Load Distribution Design Pattern those are Case Based Reasoning, Database Access and Master Slave Design Patterns. Input class will request the server for solving multi object optimization problem server will pick appropriate fitness function for the problem by using case based reasoning Design Pattern. Based on fitness function server will generate population, population of server is distributed to different clients by using Master

Slave Design Pattern. Results of different client are stored in database by using database access Design Pattern.

### 3.6. Participants:

(a) **Client:** Input Stream supplies Multi object Optimization problem to server, based on input stream problem server choose appropriate fitness function for population generation.

(b) **Server:** Server will take input from the input stream, based on the input it will find the fitness function with the help of the Case Based Reasoning Design Pattern, based on the fitness function server will find the possible chromosomes for the problem. Each chromosome is distributed different clients. After completion of the evaluation server collect results from server. Based on results server chooses appropriate solution as a result of the problem.

(c) **Client repository:** Client repository stores services that are invoked by client, repository will create separate thread for each client, if service is finished with in time stamp then it will report result to client otherwise it report service class, service class will take decision based on time stamp availability if time is available then it choose resume service otherwise it choose suspend service.

(d) **Client:** Client class creates new thread for each and every population of the server. Client uses Master Slave Design Pattern for evaluating Genetic Algorithm population.

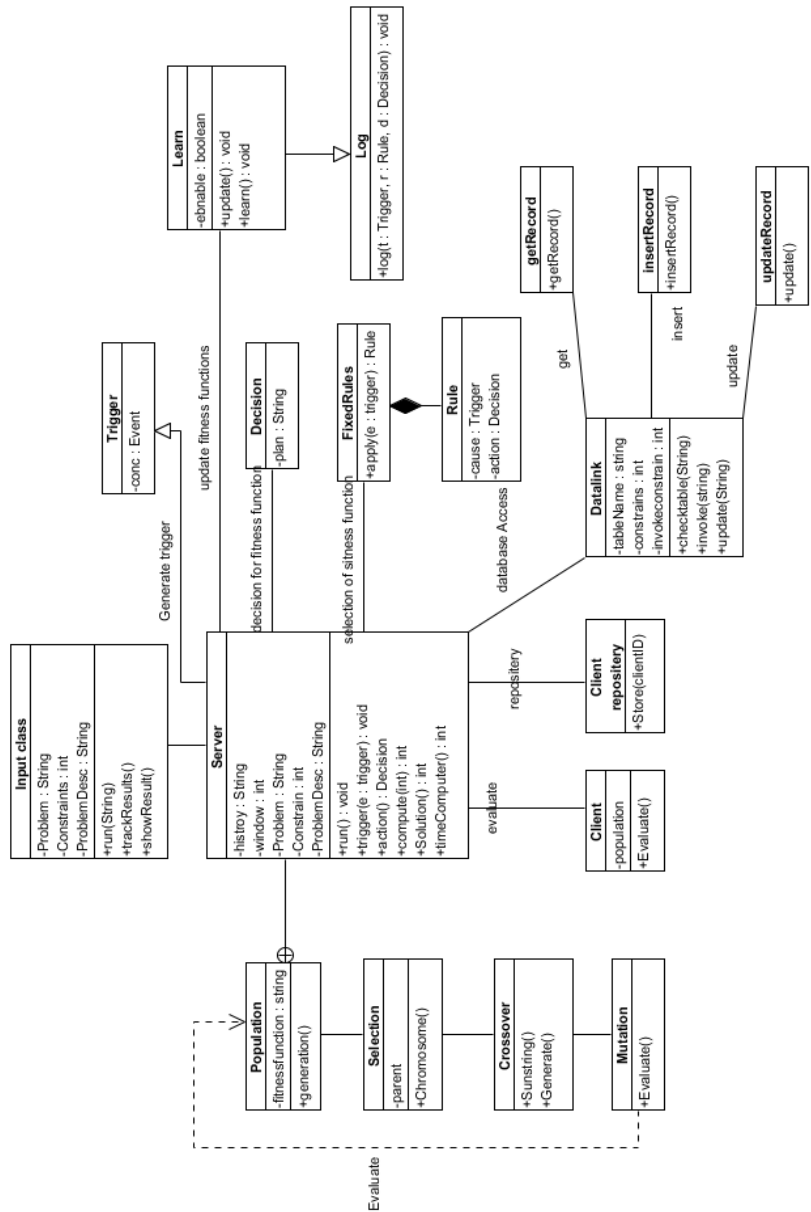

Figure 2: Class Diagram for Load Distribution Design Pattern.

(e) **Decision:** This class represents a reconfiguration plan that will yield the desired behavior in the system.

(f) **Fixed Rules:** This class contains a collection of Rules that guide the Inference Engineering producing a Decision. The individual Rules stored within the Fixed Rule scan be changed at run time.

91

(g) **Learner:** This is an optional feature of the Case based Reasoning Design Pattern.

(h) **Log:** This class is responsible for recording which reconfiguration plans have been selected during execution. Each entry is of the form Trigger-Rule-Decision.

(g) **Rule:** A Rule evaluates to true if an incoming Trigger matches the Trigger contained in the Rule.

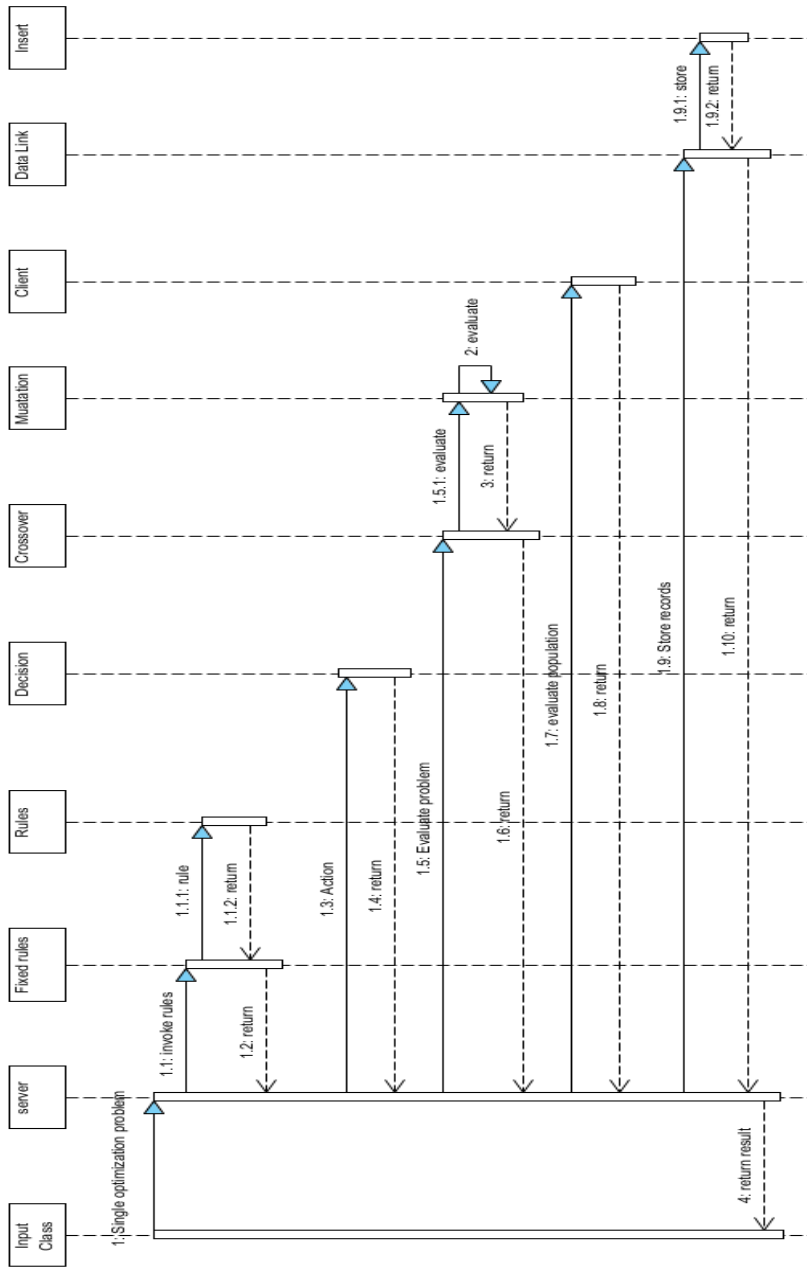

Figure 3: Sequence Diagram for Load Distribution Design Pattern.

### 3.7. Related Design Patterns:

The intent of the Service Administration pattern is similar to the Configuration pattern. The Configuration pattern decouples structural issues related to configuring services in distributed applications from the execution of the services themselves. The **Configuration pattern [1]** has been used in frameworks for configuring distributed systems to support the construction of a distributed system from a set of components. In a similar way, the Service Administration Design Pattern decouples service initialization from service processing. The primary difference is that the Configuration pattern focuses more on the active composition of a chain of related services, whereas the Service Administration Design Pattern focuses on the dynamic initialization of service handlers at a particular endpoint. In addition, the Service Administration Design Pattern focuses on decoupling service behavior from the service's concurrency strategies.

The **Manager Pattern [7]** manages a collection of objects by assuming responsibility for creating and deleting these objects. In addition, it provides an interface to allow clients access to the objects it manages. The Service Administration Design Pattern can use the Manager pattern to create and delete Services as needed, as well as to maintain a repository of the Services it creates using the Manager Pattern. However, the functionality of dynamically configuring, initializing, suspending, resuming, and terminating a Service created using the Manager Pattern must be added to fully implement the Service Administration Pattern.

### 3.9 Applicability

Use the autonomic system using Design Pattern when

- The strategy chooses the reconfiguration plan based on the input stream of the modules.
- You need different variants of an algorithm. For example, you might define algorithms reflecting different space/time trade-offs. Strategies can be used when these variants are implemented as a class hierarchy of algorithms.
- If the strategy will store different fitness functions updating the fitness function are also possible in strategy.
- It reduces the workload of server; the system is suitable for distributed computing.

## 4   SIMULATION RESULTS

A system is self-optimizing when it is capable of adapting to meet current requirements and also of taking necessary actions to self-adjust to better its performance. Resource management (e.g., load balancing) is an aspect of self-optimization. In our experiments we evaluate the e effectiveness of our pro-posed multimodal architecture. In order to make our proposal clear we have successfully developed some critical parts of our system i.e., we have developed the code for the self-optimization modules. We used Java Management Extensions (JMX)-compliant monitoring tool called JVM Monitor\footnote{http://www.jvmmonitor.org} for evaluating the self-optimization characteristic of the developed application. The simulation results for the self-optimization modules are collected with respect to:

- Used Heap memory
- Total Started Thread Count
- Process CPU Time
- Total Compilation Time

International Journal on Soft Computing (IJSC) Vol.3, No.3, August 2012

So, these are the four factors through which we have analyzed the results for the self-optimization phase with and without Design Patterns.

For self-optimization the following measures have been used:

1.  Identify the object consuming heap memory that increased during a certain duration, and identify the objects that are keeping the reference to it.
2.  Select the thread that has high load of CPU, so that from its stack traces we can find out which methods are being invoked.
3.  total number of threads
4.  peak number of threads
5.  Number of live threads.
6.  Stability: how easy or difficult is it to keep the system in a consistent state during modification [19].

We have successfully executed the application with and without applying Design Patterns and observed the following results which are given in the form of graphs as follows:

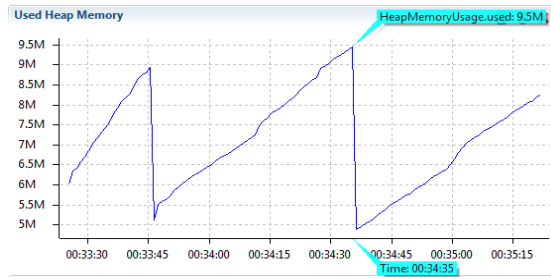

Figure 4: Heap memory usage of Self-optimization Module before applying pattern

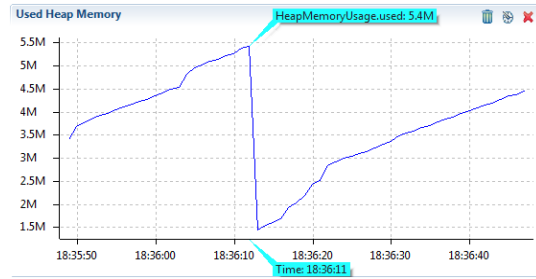

Figure 5: Heap memory usage of Self-optimization Module after applying pattern

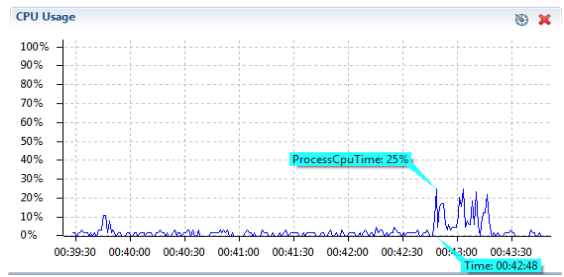

Figure 6: CPU Usage of Component Self-optimization before applying pattern

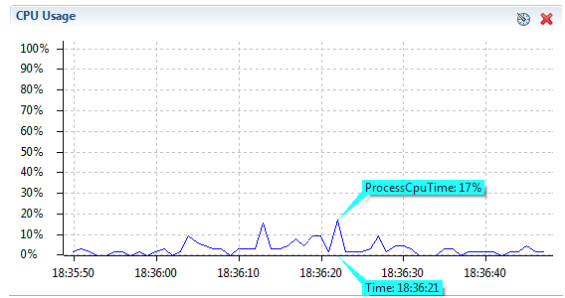

Figure 7: CPU Usage of Self-optimization Module after applying pattern

## 5 DISCUSSION

From the Figures 4 and 5 we can evaluate that the amount of Heap Memory used by applying aspectual Design Pattern is 62,345 kbytes and where as for the amount of Heap Memory used without any Design Pattern is 15,691 kbytes. It's clear that the self-optimization module with Design Pattern takes less heap memory when compared to self-optimization module without any Design Pattern.

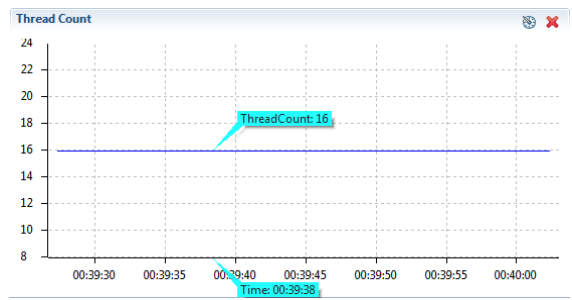

Figure 8: Thread Count of Self-optimization Module before applying pattern

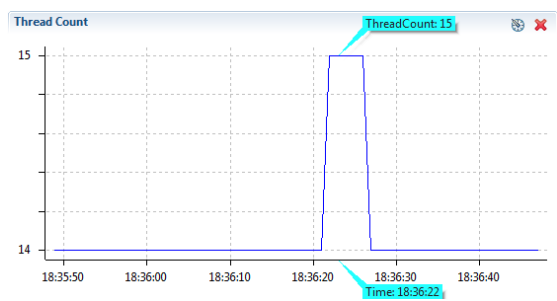

Figure 9: Thread Count of Self-optimization Module after applying pattern

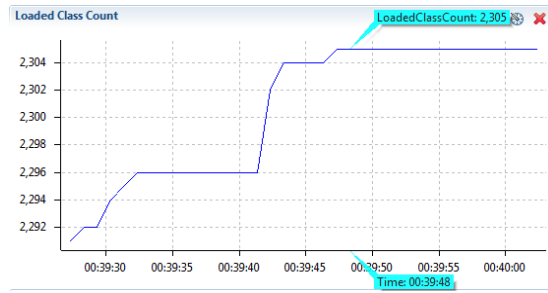

Figure 10: Loaded Class Count of Self-optimization Module before applying pattern

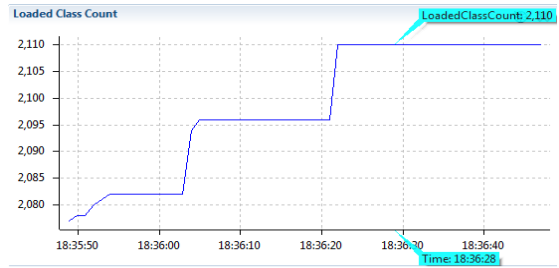

Figure 11: Loaded Class Count of Self-optimization Module after applying pattern

From the Figures 6 and 7 we can evaluate that the amount of CPU Time used by applying aspectual Design Pattern is 12.456sec and where as for the amount of CPU Time used without any Design Pattern is 16.654 sec. It's clear that the Component self-optimization module with Design Pattern takes less CPU Processing time when compared to self-optimization module without any Design Pattern.

From the Figures 8 and 9 we can evaluate that the Total Started Thread Count by applying Design Pattern is 14 and where as for Total Started Thread Count without any Design Pattern is 15. It's clear that the self-optimization module with Design Pattern takes less Total Started Thread Count when compared to self-optimization module without any Design Pattern. From Figure 10 and 11 shows loaded class count of self-optimization module before and after applying patterns.

The Total Compilation Time for the self-optimization Module with Design Pattern took 2532 sec. whereas the same Total Compilation Time for this module without any Design Pattern applied took 3587 sec. So from this we can say that even the compilation time is having good improvement when we are using Design Patterns for the implementation of the application. So from the above described comparison factors we want to make it clear that by using the Design Patterns the performance of the application/system is more e-client than without using any Design Pattern.

## 6  Conclusion

In this paper we prepare an approach for autonomic computing using the composition of Design Patterns. The system will reduce workload of server by distributing the population to different clients. This paper implies four Design Patterns those are Case Based Reasoning, Database Access Patterns and Master Slave Design Pattern. Database Access Patterns are used for storing results in database. Based on the fitness function chromosomes assign to the different clients, clients will do the job and return the result of the job finally we will evaluate results and provide best result out of all possible results.

**Future work:** Future aims to develop a system that will distribute the Genetic Algorithm to different clients, and design an autonomic system that will reconfigure based on autonomic changes in the system using Design Patterns.

## A. Interfaces Definition for the Composite Design Pattern Entities

**Input Class:**
```
Public class Inputclass
{
   public String Read()
   {}
}
```

**Server:**
```
Public class Server
{
      Public string fitnessfunction()
      {
      i.concereteImpl();
      }
      Public int enqueuejob(int )
      {}
      Public int jobqueue(int)
      {}
      Public int result()
      {}
      Public getjob()
      {}
}
```
**Client:**
```
Public class Client
{
      Public int dojob()
   {}
}
```
**Insersion:**
```
 Public class Viewone
{
    Public int insert(object e)
    {}
}
```
**Fixed rules:**
```
Public class Fixedrules
```

97

```
{
   Public void Elaborate(String str)
   { }
}
Public class rules implements Fixedrules
{
   Public void Elaborate(String str)
   { }
 }
```

**Decision:**

```
Public class Decision
{
    Public int action(object )
     { }
}
```

**Client repository:**

```
class Clientrepositery
 {
     private static Client instance = null;
     public static instance()
    {
      if( instance == null )
      {
        instance = new Singleton();
      }
      return instance;
    }
```

International Journal on Soft Computing (IJSC) Vol.3, No.3, August 2012